\documentclass{sig-alternate}

\usepackage{graphicx}
\usepackage{tabularx}
\usepackage{amsopn}
\usepackage[ruled,linesnumbered]{algorithm2e}
\usepackage{bm}
\usepackage{booktabs}
\usepackage{epsfig}
\usepackage{dcolumn}
\usepackage{bm}
\usepackage{subfigure}
\usepackage{mathtools}
\usepackage{amssymb}
\usepackage{amsfonts}
\usepackage{enumerate}
\usepackage{comment}
\usepackage{mathrsfs} 
\usepackage{flushend}
\usepackage{cuted}
\usepackage{cite}
\usepackage{mathrsfs} 

\begin{document}

\title{Detecting Communities in Tripartite Hypergraphs}

\numberofauthors{2}
\author{
\alignauthor Xin Liu\\
       \affaddr{Tokyo Institute of Technology}\\
       \affaddr{W8-59 2-12-1 Ookayama, Meguro}\\
       \affaddr{Tokyo, 152-8552 Japan}\\
       \email{tsinllew@ai.cs.titech.ac.jp}
\alignauthor Tsuyoshi Murata\\
       \affaddr{Tokyo Institute of Technology}\\
       \affaddr{W8-59 2-12-1 Ookayama, Meguro}\\
       \affaddr{Tokyo, 152-8552 Japan}\\
       \email{murata@cs.titech.ac.jp}
}

\maketitle

\section{Introduction}\label{sec1}
In social tagging systems, users collaboratively manage tags to annotate resources. Naturally, social tagging systems can be modeled as a tripartite hypergraph, where there are three different types of nodes, namely users, resources and tags, and each hyperedge has three end nodes, connecting a user, a resource and a tag that the user employs to annotate the resource.

\begin{figure}[!t]
\centering
\subfigure[][]{\label{fig2a}
\includegraphics[width=0.20\textwidth, bb= -30 15 400 320]{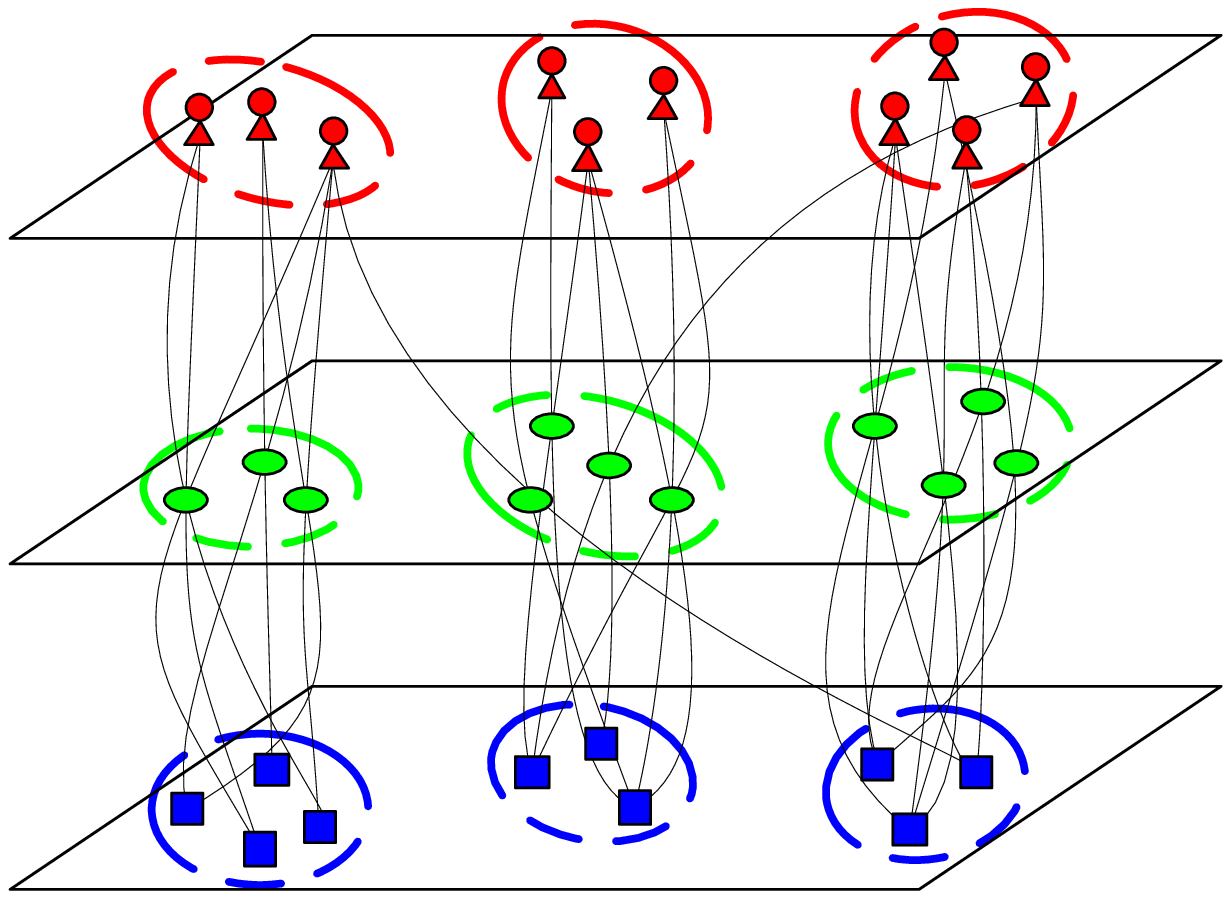}
}
\subfigure[][]{\label{fig2b}
\includegraphics[width=0.20\textwidth, bb= -30 15 400 320]{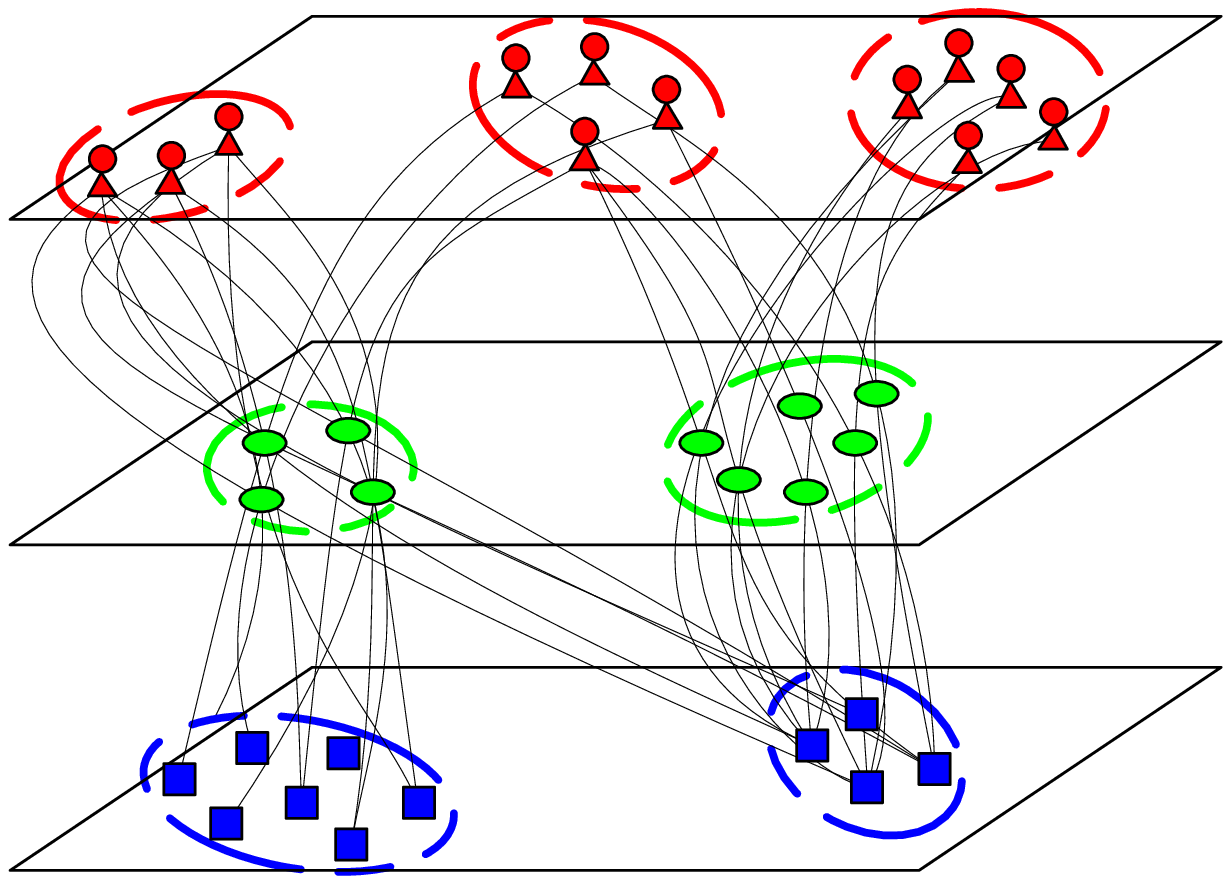}
}
\caption{\label{fig2} Communities with (a) one-to-one correspondence, and (b) many-to-many correspondence.}
\end{figure}

As for community detection in tripartite hypergraphs, a common strategy is to reduce the tripartite hypergraph to simpler unipartite graphs, bipartite graphs, or tripartite graphs, and then detect communities in the corresponding graphs \cite{ZlaticHypergraphQuantities,NeubauerKPartiteCommunity,LuHyperedgeDecomposition}. One major drawback of this class of methods is that some valuable information of the original hyperedges is lost during reduction \cite{ZhouBipartiteNetProjection}, and the subsequently detected communities tend to be less accurate. Researchers also proposed extended modularity optimization \cite{MurataTriModularityWWW} and tensor decomposition \cite{LinMetaFac} methods. But these two methods are biased towards communities with one-to-one correspondence, as shown in Fig~\ref{fig2a}. Real-world social tagging systems are often more complex than that. For example, a group of users may be interested in a collection of resources about programming technology and another collection about sports. Hence, communities with many-to-many correspondence, as shown in Fig~\ref{fig2b}, are more significant. Besides, another disadvantage of some previous methods \cite{LuHyperedgeDecomposition,LinMetaFac} is that they require one to specify certain parameters such as the numbers of communities. In practice, such a priori knowledge is difficult to obtain.

In this paper, we propose a quality function, based on the minimum description length (MDL) principle \cite{RissanenMDL}, for measuring the goodness of different partitions of a tripartite hypergraph into communities, and develop a community detection algorithm based on minimizing the quality function. Our method overcomes the limitations of previous methods and has the following key properties:

\begin{small}
\begin{itemize}
\itemsep=-1.0pt      
\item Independent: it handles broad families of tripartite hypergraphs, and is competent for both communities with one-to-one correspondence and many-to-many correspondence.
\item Parameter-free: given the structure of a tripartite hypergraph, it can automatically detect communities, without any prior knowledge like the numbers of communities.
\item Accurate: it is more accurate than previous methods.
\item Scalable: it is fast and scalable to large-scale hypergraphs.
\end{itemize}
\end{small}

\section{Problem Formulation}\label{sec3}

One fundamental issue is the definition of community in tripartite hypergraphs. Generally, a community should be a group of related nodes that correspond to a functional subunit in the real-world system. In unipartite graphs, a community is often understood as a group of nodes with dense connections between them. But this notion is not suitable for tripartite hypergraphs, since nodes of the same type are not connected. Instead, we consider a tripartite hypergraph community as a group of nodes that are structurally equivalent (in a weakened sense, and the same hereinafter) \cite{NewmanNetworksAnIntroduction}. This is a natural assumption, because a group of nodes that are similar with one another as regards their relations to nodes of other types are very likely to form a functional subunit. For example, in a social tagging system, those users having similar tagging actions are very likely to share the same interests; those resources that are annotated with similar tags are very likely to be in the same category. Meanwhile, the dense connections between certain communities in different node sets constitute their correspondence.

In the following, we formulates the problem of community detection in tripartite hypergraphs. Now assume an undirected and unweighted tripartite hypergraph $\bm{H} = ({\bm{V}^r} \cup {\bm{V}^g} \cup \bm{V}^b, \bm{E})$, where $\bm{V}^r$, $\bm{V}^g$ and $\bm{V}^b$ are three disjoint node sets, and $\bm{E} \subseteq \{( v_i^r \in \bm{V}^r,  v_j^g\in \bm{V}^g, v_k^b \in \bm{V}^b)\}$ is the set of three-way hyperedges. For simplicity, nodes of the three different types $v_i^r \in \bm{V}^r$, $v_j^g\in \bm{V}^g$ and $v_k^b \in \bm{V}^b$ are colored red, green and blue, respectively. Suppose $n^r = |\bm{V}^r|, n^g = |\bm{V}^g|$ and $n^b = |\bm{V}^b|$ are the numbers of red, green and blue nodes, and $m = |\bm{E}|$ the number of hyperedges. The structure of $\bm{H}$ can be represented by a three-dimensional binary array $\mathbf{A}$ of $n^r \times n^g \times n^b$ size, with elements
\begin{equation*} A_{ijk}=
\begin{cases}
1 & \text{if }(v_i^r,v_j^g,v_k^b)\in\bm{E}; \\
0 & \text{otherwise}.
\end{cases}
\end{equation*}

\begin{table}[!t]
\centering    
\caption{\label{table1}Notations for a tripartite hypergraph $\bm{H}$}
\begin{scriptsize}   
\begin{tabular}{ll}
\toprule
\textrm{Symbol}&\textrm{Meaning}\\
\midrule
$\bm{V}^r$ & The red node set \\
$\bm{V}^g$ & The green node set \\
$\bm{V}^b$ & The blue node set \\
$\bm{E}$ & The hyperedge set \\
$n^r$ & The number of red nodes \\
$n^g$ & The number of green nodes \\
$n^b$ & The number of blue nodes \\
$m$   & The number of hyperedges \\
$v_i^r$ & The $i$-th red node \\
$v_j^g$ & The $j$-th green node \\
$v_k^b$ & The $k$-th blue node \\
$c^r$ & The number of red communities \\
$c^g$ & The number of green communities \\
$c^b$ & The number of blue communities \\
$\bm{V}_{\alpha}^r$ & The $\alpha$-th red community \\
$\bm{V}_{\beta}^g$ & The $\beta$-th green community \\
$\bm{V}_{\gamma}^b$ & The $\gamma$-th blue community \\
$n_{\alpha}^r$ & The number of nodes in $\bm{V}_{\alpha}^r$\\
$n_{\beta}^g$ & The number of nodes in $\bm{V}_{\beta}^g$\\
$n_{\gamma}^b$ & The number of nodes in $\bm{V}_{\gamma}^b$\\
$r_{i}$ & The vector $\mathbf{r}$ whose $i$-th element indicates\\
        & the community membership of $v_i^r$ \\
$g_{j}$ & The vector $\mathbf{g}$ whose $j$-th element indicates\\
        & the community membership of $v_j^g$ \\
$b_{k}$ & The vector $\mathbf{b}$ whose $k$-th element indicates\\
        & the community membership of $v_k^b$ \\
$A_{ijk}$ & The three-dimensional array $\mathbf{A}$ whose $(i,j,k)$\\
          & element indicates the number of hyperedges\\
          & between $v_i^r$, $v_j^g$ and $v_k^b$ \\
$M_{\alpha\beta\gamma}$ & The three-dimensional array $\mathbf{M}$ whose $(\alpha,\beta,\gamma)$ \\
                        & element indicates the number of hyperedges\\
                        & between $\bm{V}_{\alpha}^r$, $\bm{V}_{\beta}^g$ and $\bm{V}_{\gamma}^b$\\
\bottomrule
\end{tabular}
\end{scriptsize}
\end{table}

The problem of community detection in $\bm{H}$ is that, given $\mathbf{A}$, how can we find a good partition $\bm{\mathscr{C}}=\{\bm{V}_{\alpha}^r\}_{\alpha=1}^{c^r} \oplus \{\bm{V}_{\beta}^g\}_{\beta=1}^{c^g} \oplus \{\bm{V}_{\gamma}^b\}_{\gamma=1}^{c^b}$ that divides $\bm{V}^r$, $\bm{V}^g$ and $\bm{V}^b$ into disjoint communities, respectively:
\begin{eqnarray*}
\oplus \{\bm{V}_{\alpha}^r\}_{\alpha=1}^{c^r}&=&\bm{V}^r\\
\oplus \{\bm{V}_{\beta}^g\}_{\beta=1}^{c^g}&=&\bm{V}^g\\
\oplus \{\bm{V}_{\gamma}^b\}_{\gamma=1}^{c^b}&=&\bm{V}^b
\end{eqnarray*}
Note that the numbers of red, green and blue communities $c^r$, $c^g$ and $c^b$ are not known a priori. The meaning of ``good'' is twofold: i) nodes in the same community are structurally equivalent; ii) hyperedges between communities are either dense or sparse, so that the correspondence between communities is clear.

Throughout the paper, we use Latin letters $i$, $j$ and $k$ for indices of red, green and blue nodes, respectively, and use Greek letters $\alpha$, $\beta$ and $\gamma$ for indices of red, green and blue communities, respectively. Table~\ref{table1} summarizes the notations used in this paper.

\section{The Proposed Method}\label{sec4}

In this section, we first define a quality function for measuring the goodness of different partitions of a tripartite hypergraph into communities, and then propose an algorithm for minimizing the quality function.

When we describe a graph as a set of communities, we are highlighting certain regularities (e.g., the similarities of nodes in the same community and the dissimilarities of nodes between different communities) while filtering out relatively unimportant details (e.g., the dissimilarities of nodes in the same community). Thus, description of a graph as communities can be viewed as a lossy compression of that graph's structure, and the community detection problem as a problem of finding an efficient compression of the structure. This is the main insight of the structural information compression method proposed in \cite{RosvallInfoCompression}, where the authors focus on information compression on a unipartite graph's structure. Here we show how to compress the structural information of a tripartite hypergraph, in order to formulate our quality function.

Now let us envision a communication process of transmitting structural information of a tripartite hypergraph $\bm{H}$. A signaler knows the structure of $\bm{H}$ and aims to transmit much of the information in a reduced fashion to a receiver over a noiseless channel. To do so, the signaler makes a partition of $\bm{H}$ into communities and encodes the structural information X = \{$\mathbf{A}$\} as compressed information summarizing the community structure: Y = \{$\mathbf{r, g, b, M}$\}, where $\mathbf{r}$, $\mathbf{g}$ and $\mathbf{b}$ are the community membership vectors of red, green and blue nodes, and $\mathbf{M}$ is the community connectivity array. For a partition dividing red, green and blue nodes into $c^r$, $c^g$ and $c^b$ communities, we have $\mathbf{r}=[r_1,r_2,\ldots, r_{n^r}]$, $\mathbf{g}=[g_1,g_2,\ldots, g_{n^g}]$ and $\mathbf{b}=[b_1,b_2,\ldots, b_{n^b}]$, where $r_i \in \{1, 2, \ldots, c^r\}$, $g_j \in \{1, 2, \ldots, c^g\}$ and $b_k \in \{1, 2, \ldots, c^b\}$ indicate the community memberships of nodes $v_i^r$, $v_j^g$ and $v_k^b$, respectively. The community connectivity array $\mathbf{M}$ is a three-dimensional array of $c^r \times c^g \times c^b$ size, with element $M_{\alpha\beta\gamma}\in\{0,1,2,\ldots,m\}$ indicating the number of hyperedges between communities $\bm{V}_{\alpha}^r$, $\bm{V}_{\beta}^g$ and $\bm{V}_{\gamma}^b$. That is
\begin{eqnarray*}
M_{\alpha\beta\gamma}=\sum_{v_i^r \in \bm{V}_{\alpha}^r}\sum_{v_j^g \in \bm{V}_{\beta}^g}\sum_{v_k^b \in \bm{V}_{\gamma}^b}{A_{ijk}}
\end{eqnarray*}
It is easy to derive that the description length (in bits) of the compressed information Y is
\begin{eqnarray*}
\text{L(Y)}=n^r\mathbf{\log}c^r+n^g\mathbf{\log}c^g+n^b\mathbf{\log}c^b+c^rc^gc^b\mathbf{\log}(m+1)
\end{eqnarray*}
where the logarithm is taken in base 2.

After receiving Y, the receiver knows the community membership of each node and the number of hyperedges between each community triple. Then he tries to recover the original structural information X by constructing possible candidates. The number of different candidates is given by
\begin{eqnarray*}
\prod_{\alpha=1}^{c^r} \prod_{\beta=1}^{c^g} \prod_{\gamma=1}^{c^b} {n_{\alpha}^r n_{\beta}^g n_{\gamma}^b \choose M_{\alpha\beta\gamma}}
\end{eqnarray*}
where $n_{\alpha}^r=|\bm{V}_{\alpha}^r|$, $n_{\beta}^g=|\bm{V}_{\beta}^g|$ and $n_{\gamma}^b=|\bm{V}_{\gamma}^b|$ are the numbers of nodes in communities $\bm{V}_{\alpha}^r$, $\bm{V}_{\beta}^g$ and $\bm{V}_{\gamma}^b$, the parentheses denote the binomial coefficient, and each binomial coefficient gives the number of different candidates for recovering the original $M_{\alpha\beta\gamma}$ hyperedges between $\bm{V}_{\alpha}^r$, $\bm{V}_{\beta}^g$ and $\bm{V}_{\gamma}^b$. Hence, the description length of the additional information for the receiver to recover X (i.e. the conditional information between X and Y) is
\begin{eqnarray*}
\text{L(X$|$Y)}=\log \left[ \prod_{\alpha=1}^{c^r} \prod_{\beta=1}^{c^g} \prod_{\gamma=1}^{c^b} {n_{\alpha}^r n_{\beta}^g n_{\gamma}^b \choose M_{\alpha\beta\gamma}} \right].
\end{eqnarray*}

The objective is that signaler transmits the least while the receiver receives the most. Intuitively, if the signaler makes a ``good'' partition as described in Section~\ref{sec3}, which capitalizes on regularities in the hypergraph's structure, the compression based on it would achieve the optimal trade-off between L(Y) and L(X$|$Y). According to the minimum description length (MDL) principle \cite{RissanenMDL},
\begin{align}
&\text{Q}(\bm{\mathscr{C}})\nonumber=\text{L(Y)}+\text{L(X$|$Y)}\nonumber\\
\begin{split}
    &=n^r\mathbf{\log}c^r+n^g\mathbf{\log}c^g+n^b\mathbf{\log}c^b+c^rc^gc^b\mathbf{\log}(m+1)\\
    & \quad +\log \Bigg[ \prod_{\alpha=1}^{c^r} \prod_{\beta=1}^{c^g} \prod_{\gamma=1}^{c^b} {|\bm{V}_{\alpha}^r||\bm{V}_{\beta}^g||\bm{V}_{\gamma}^b| \choose \sum\limits_{v_i^r \in \bm{V}_{\alpha}^r} \sum\limits_{v_j^g \in \bm{V}_{\beta}^g} \sum\limits_{v_k^b \in \bm{V}_{\gamma}^b}{A_{ijk}}} \Bigg]
\end{split}\label{eq1}
\end{align}
would get the minimum value. This is the quality function for measuring the goodness of a partition $\bm{\mathscr{C}}$ of a tripartite hypergraph into communities.

Now we can evaluate a partition based on the quality function Q, and a low value of Q indicates a good partition. So the task is to search over all possible partitions for one that has a minimum Q. However, like modularity optimization, finding the global optimal solution is NP-hard \cite{BrandesModularityNPCompleteness}. We develop an approximate algorithm that can be implemented in near linear time, as presented in Algorithm~\ref{alg1}. For more information, please refer to \cite{LiuLPAmplus, BlondelFastUnfolding}.

\SetAlFnt{\scriptsize}
\begin{algorithm}[!t]
\DontPrintSemicolon
\KwIn{Connectivity array $\scriptsize{\mathbf{A}}$ of $\scriptsize{\bm{H}}$}
\KwOut{Partition of $\scriptsize{\bm{H}}$ into communities}
\Begin{
    \tcp{\tiny{Phase 1}}
    assign each node in $\bm{H}$ a unique label;\;
    \Repeat{a local minimum of \textup{Q}}{
            update each node's label;\;
    }

    \BlankLine
    \Repeat{no change in \textup{Q}}{
        \tcp{\tiny{Phase 2}}
        build a reduced tripartite hypergraph $\bm{H'}$;\;
        assign each node in $\bm{H'}$ a unique label;\;
        \Repeat{a local minimum of \textup{Q}}{
            update each node's label;\;
        }
        \BlankLine
        \tcp{\tiny{Phase 1}}
        retrieve labels in $\bm{H}$ from the corresponding labels in $\bm{H'}$;\;
        \Repeat{a local minimum of \textup{Q}}{
            update each node's label;\;
        }
    }
    identity communities as groups of nodes bearing the same labels;\;
}
\caption{Detecting communities in a tripartite hypergraph $\scriptsize{\bm{H}}$ by minimizing quality function Q}\label{alg1}
\end{algorithm}

\section{Evaluation}\label{sec5}

In this section, we concentrate on comparing our method with previous ones in terms of accuracy. The basic scheme is as follows: we apply various methods to a set of synthetic tripartite hypergraphs with known community structure (the true partition), and compare the similarities between partitions obtained by different methods and the true partition; the closer of an obtained partition to the true partition, the better of the corresponding method. To quantify the similarity between two partitions, we use \textit{normalized mutual information} (NMI) \cite{DanonCompareCommunityAlgo}, which has a maximum value of 1 if two partitions match completely, and a minimum value of 0 if they are totally independent of one another.

\begin{figure*}[!t]
\centering
\includegraphics[width=0.28\textwidth, bb= 110 280 500 640]{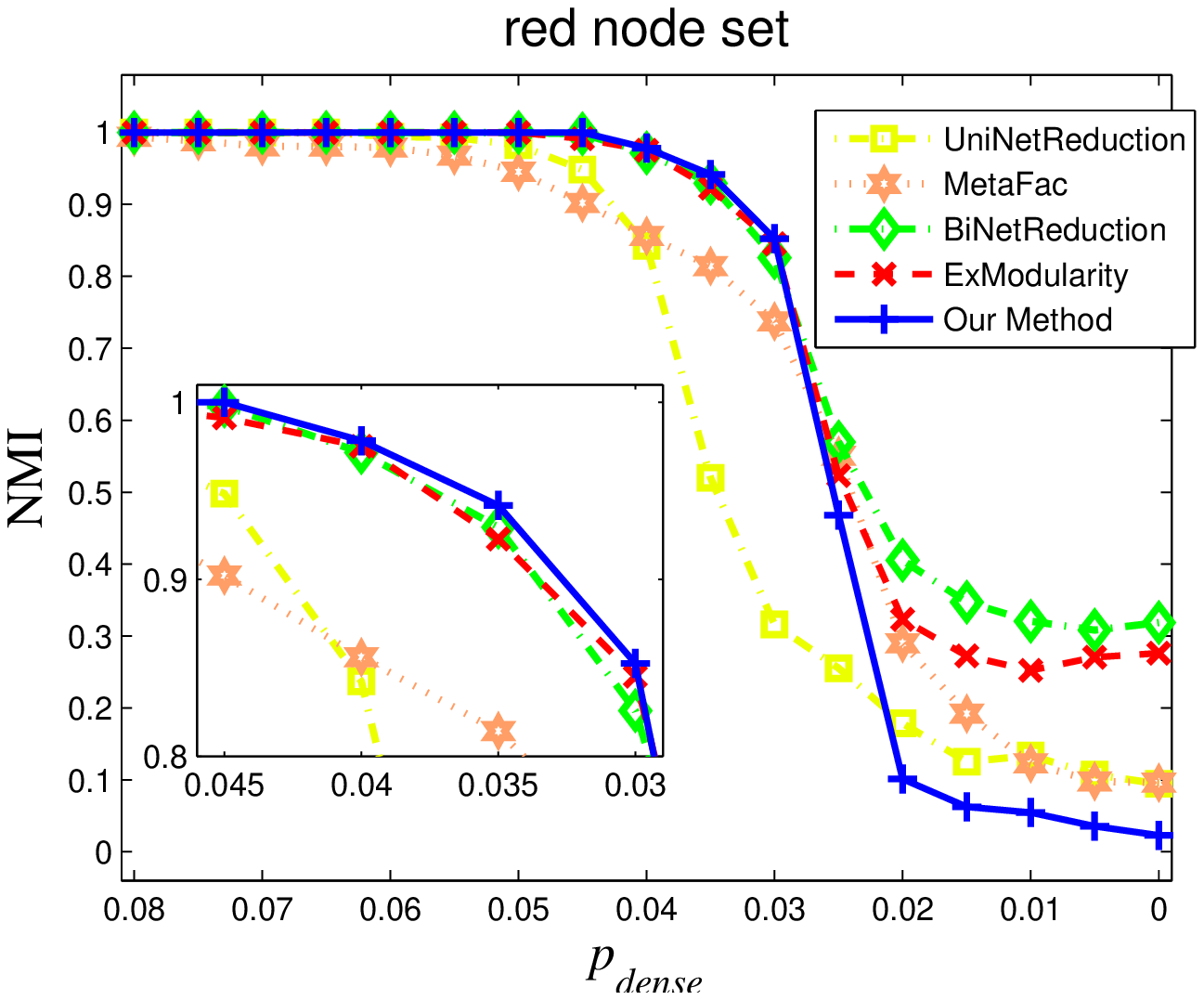}
\includegraphics[width=0.28\textwidth, bb= 110 280 500 640]{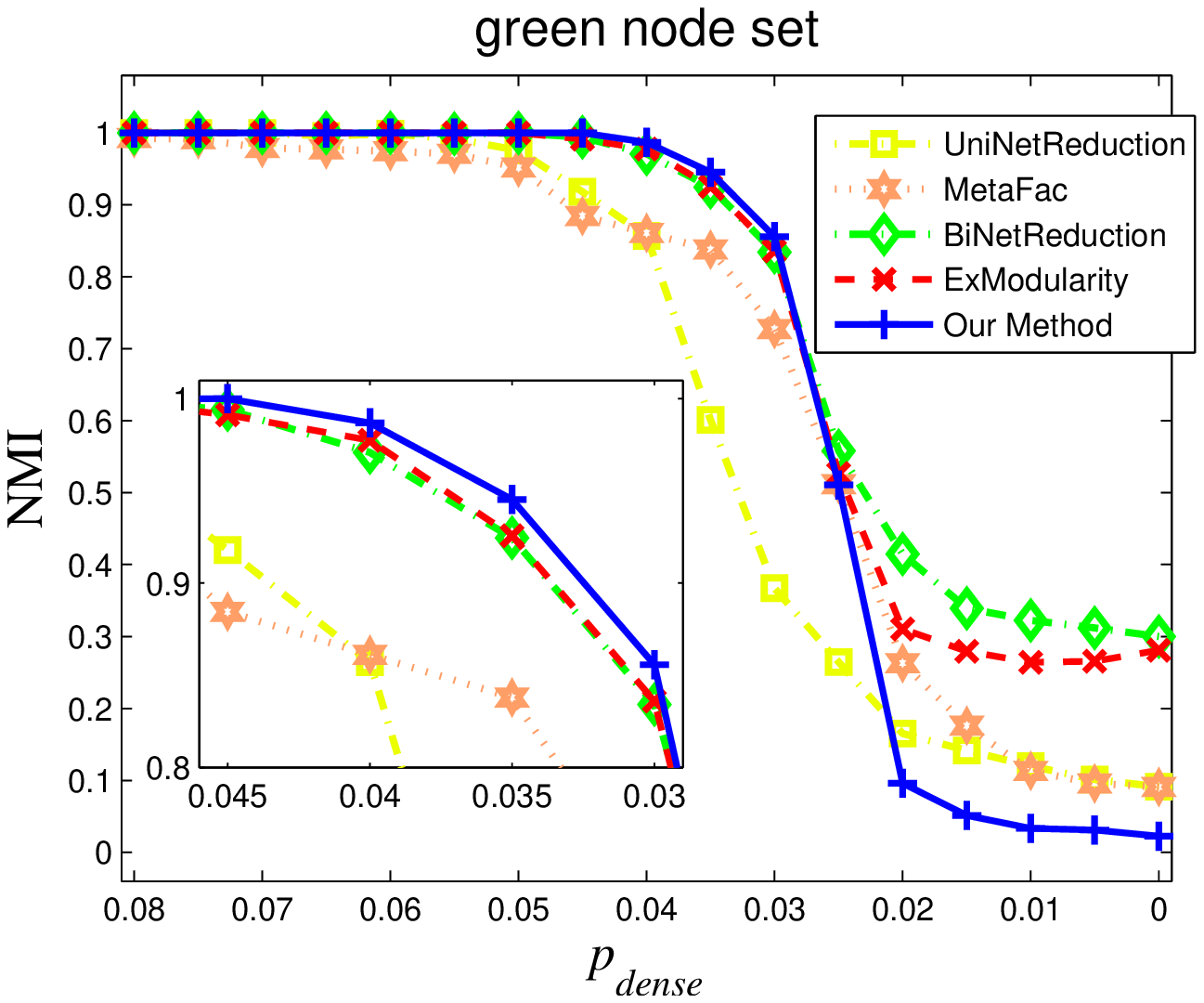}
\includegraphics[width=0.28\textwidth, bb= 110 280 500 640]{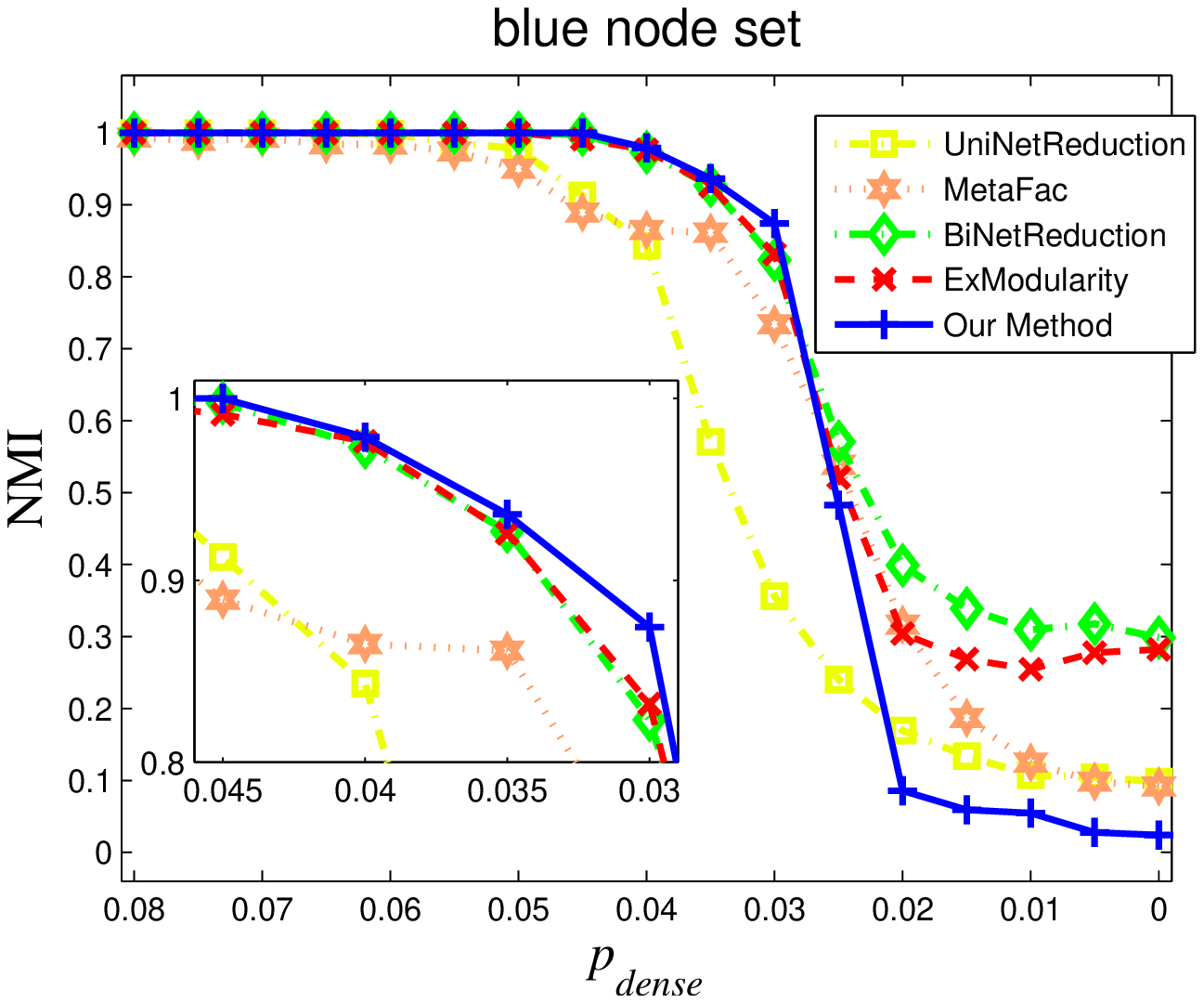}
\caption{\label{fig4}Performances in the synthetic dataset with built in communities of one-to-one correspondence.}
\end{figure*}

We consider several opponent methods that cover state of the art techniques. They are, in order, the extended modularity optimization method (ExModularity) proposed by Murata \cite{MurataTriModularityWWW}, the tensor decomposition method (MetaFac) presented by Lin et al. \cite{LinMetaFac}, and the method advanced by Neubauer et al., which involves reduction of a tripartite hypergraph to bipartite graphs (BiNetReduction) \cite{NeubauerKPartiteCommunity}. In addition, we consider another method (UniNetReduction) modified from Zlati\'c's approach \cite{ZlaticHypergraphQuantities}, which involves reduction of a tripartite hypergraph to unipartite graphs.

\begin{figure*}[!t]
\centering
\includegraphics[width=0.28\textwidth, bb= 110 285 500 640]{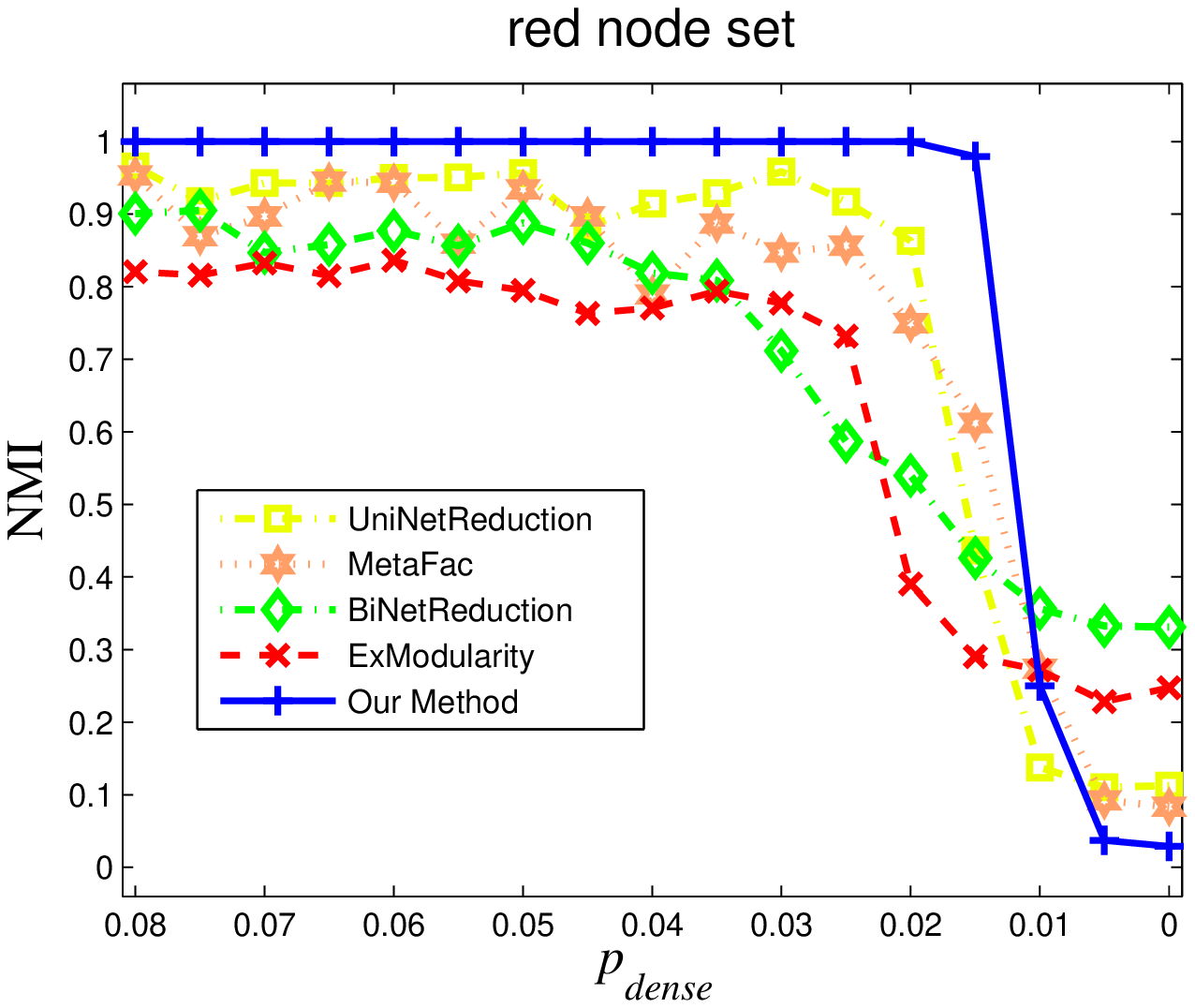}
\includegraphics[width=0.28\textwidth, bb= 110 285 500 640]{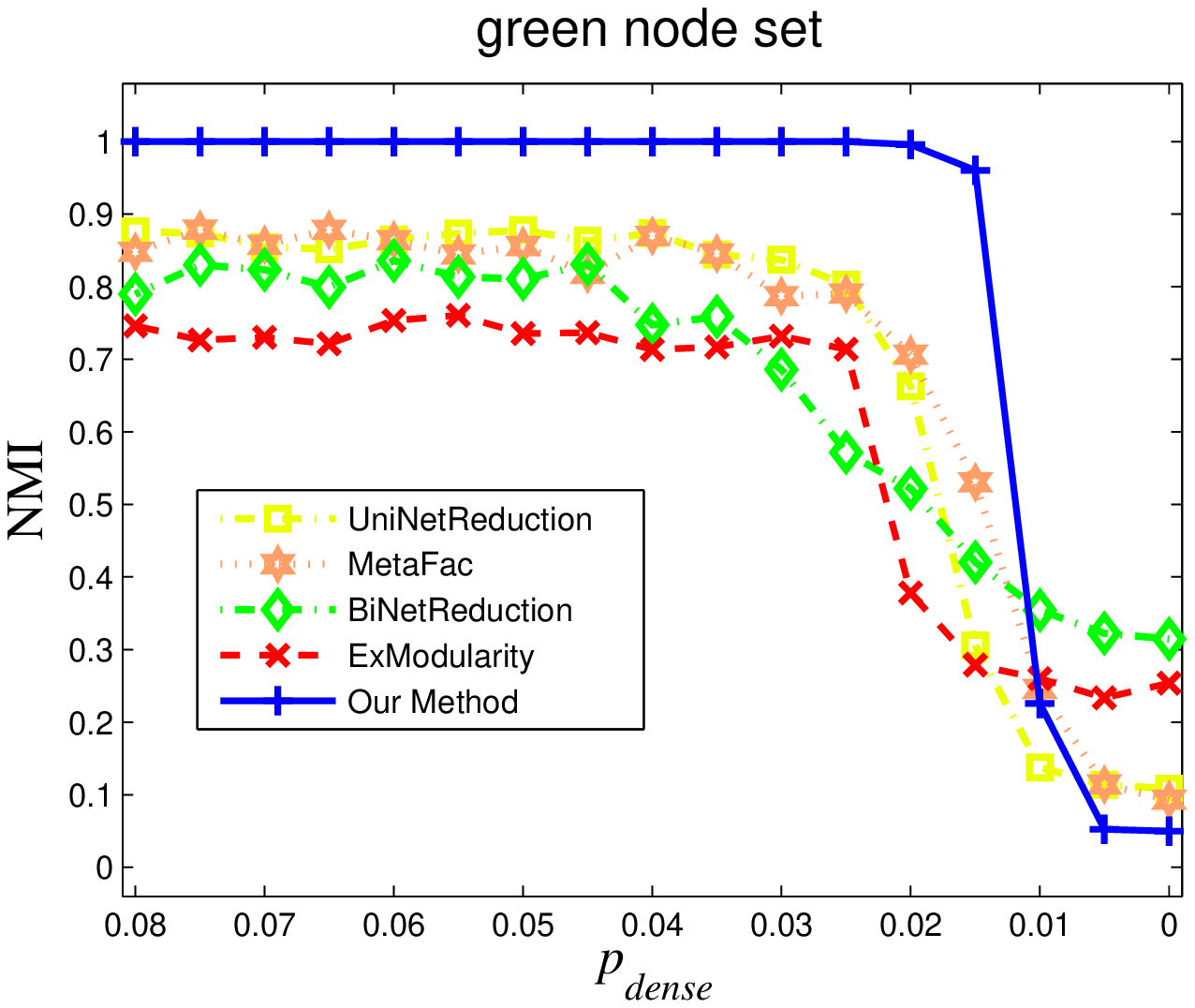}
\includegraphics[width=0.28\textwidth, bb= 110 285 500 640]{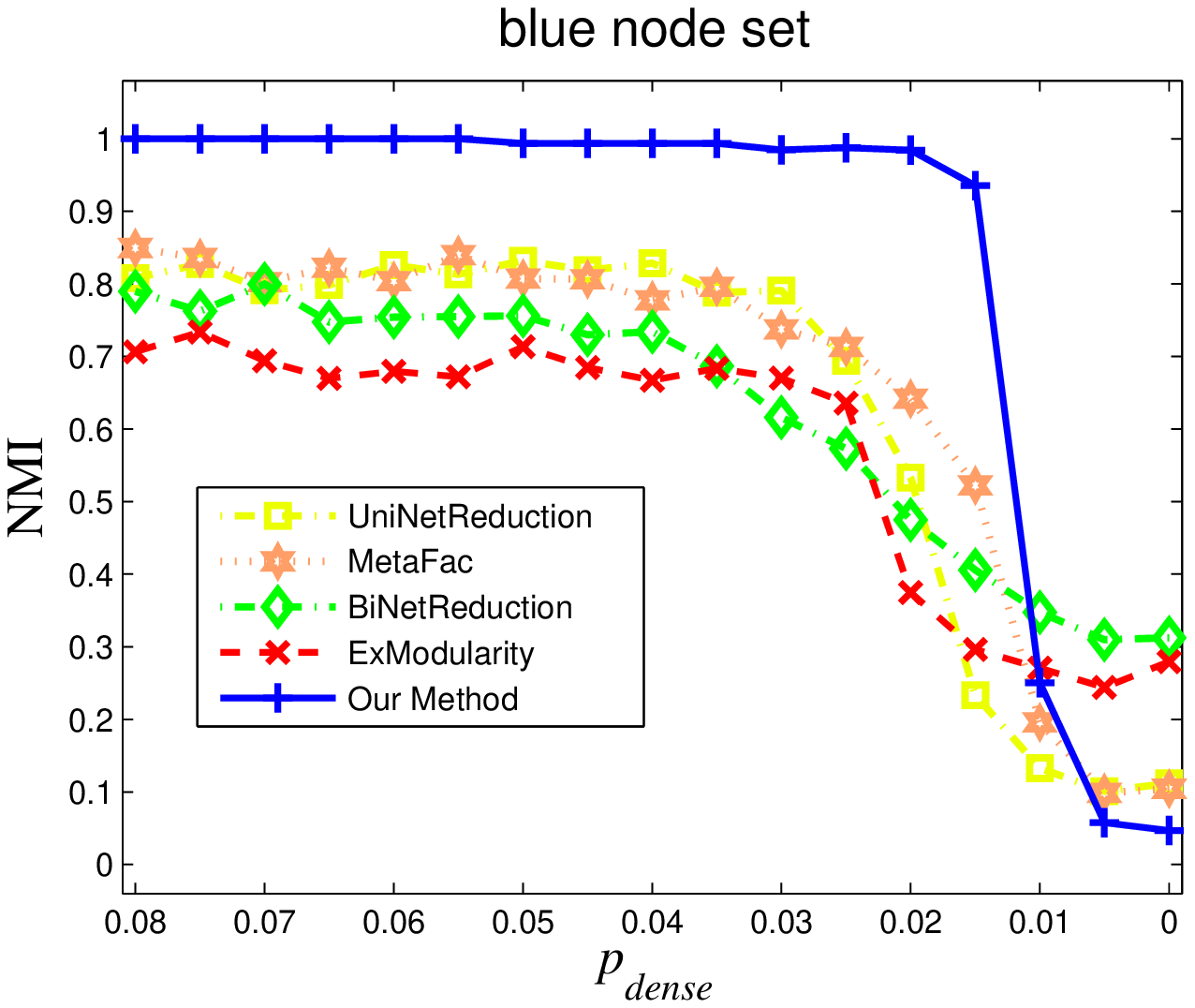}
\caption{\label{fig5}Performances in the synthetic dataset with built in communities of many-to-many correspondence.}
\end{figure*}

For the first case, we consider comparing these methods in a set of synthetic tripartite hypergraphs with built in communities of one-to-one correspondence (detailed procedure for generating the dataset is omitted here). In Fig.~\ref{fig4}, we show the performances of various methods in this set of hypergraphs. On the whole, performance of each method varies in a similar way across red, green and blue node sets (since red, green and blue nodes are in a symmetric status in the hypergraph generation procedures). Specifically, ExModularity, BiNetReduction and our method perform excellently, correctly detecting not only the numbers of communities but also community membership of each node almost all the way to the point $p_{dense}$=0.05. At the turning stage, i.e. $p_{dense}$ falling from 0.045 to 0.03, our method slightly outperforms ExModularity and BiNetReduction, as shown in the embedded figures. Thereafter, performances of the three methods deteriorate markedly. MetaFac, though given a prior knowledge of the true numbers of communities, does not provide remarkable result. The record for UniNetReduction is even worse. Its performance decreases as early as $p_{dense}$=0.055. When $p_{dense}$$\leq$0.3, it loses most of the information about the true partition.

For the second case, we generated a set of synthetic tripartite hypergraphs with built in communities of many-to-many correspondence (detailed procedure for generating the dataset is omitted here). Applying different methods to this set of synthetic hypergraphs, we calculate NMI between obtained partitions and the true partition. The results are shown in Fig.~\ref{fig5} (values are averaged over 20 runs). As Fig.~\ref{fig5} shows, our method outperforms others by a large margin. It works almost perfectly all the way until $p_{dense}$=0.015, with a sudden dramatic fall thereafter. As for other methods, we can observe three common features. 1) None of them can detect community membership with 100\% accuracy, even when $p_{dense}$=0.08. 2) Their best performances are in red node set, the middle in green node set, and the worst in blue node set. 3) Their performances deteriorate much earlier than our method, often with records fluctuating wildly before the turning points. In specific, UniNetReduction is the best among them in most of the time, followed by MetaFac. Note that MetaFac is given at least an estimate of the true numbers of communities, so its performance is not appealing. Contrary to the excellent performance in the previous set of hypergraphs, BiNetReduction and ExModularity do not show satisfactory result this time.

\section{Conclusion}\label{sec7}
Based on the information compression idea, we define a quality function for measuring the goodness of different partitions of a tripartite hypergraph into communities, and develop an algorithm for minimizing this quality function. Compared with previous methods, our method is competent for both communities with one-to-one correspondence and many-to-many correspondence. It should be emphasized that our method is parameter-free. In the future, we would like to apply our method to real-world social tagging systems.

\bibliographystyle{abbrv}
\bibliography{myRef8}

\flushend

\end{document}